\documentclass[twocolumn, twocolappendix, tighten]{aastex701}
\usepackage{amsfonts}
\usepackage{amsmath}
\usepackage{amssymb}
\usepackage{physics}
\usepackage{xcolor}

\newcommand{\E}{\mathcal{E}}
\newcommand{\AM}{\mathcal{L}_z}
\newcommand{\Q}{\mathcal{Q}}
\newcommand{\K}{\mathcal{K}}

\newcommand{\LL}[1]{\textcolor{black}{#1}}

\begin{document}

\submitjournal{ApJL}

\shorttitle{Gravitational-Wave Signatures of QPEs}
\shortauthors{Lui et al.}

\title{Gravitational-Wave Signatures of Quasiperiodic Eruptions: \\LISA Detection Prospects for RX J1301.9+2747}

\correspondingauthor{Lixin Dai \\
lixindai@hku.hk}

\author{Leif Lui}
\affiliation{Department of Physics, The University of Hong Kong, Pokfulam Road, Hong Kong}
\affiliation{Beijing Institute of Mathematical Sciences and Applications, Beijing 101408, China}
\email{}
\author{Alejandro Torres-Orjuela}
\affiliation{Beijing Institute of Mathematical Sciences and Applications, Beijing 101408, China}
\email{}

\author[orcid=0000-0003-2694-933X,sname='Kar Chowdhury', gname='Rudrani']{Rudrani Kar Chowdhury}
\affiliation{Tata Institute of Fundamental Research, Homi Bhabha Road, Mumbai 400005, India}
\affiliation{Department of Physics, The University of Hong Kong, Pokfulam Road, Hong Kong}
\email{rudrani.chowdhury@tifr.res.in}  

\author{Lixin Dai}
\affiliation{Department of Physics, The University of Hong Kong, Pokfulam Road, Hong Kong}
\affiliation{The Hong Kong Institute for Astronomy and Astrophysics, The University of Hong Kong, Pokfulam Road, Hong Kong, 
China}
\email[Corresponding author: ]{lixindai@hku.hk}

\begin{abstract}
Quasiperiodic eruptions (QPEs) are intense, recurring outbursts of X-ray radiation originating from the nuclei of distant galaxies. One of the promising models of QPE explains these eruptions using extreme mass-ratio inspirals (EMRIs), in which a stellar-mass object—such as a star or a stellar-mass black hole—orbits a central massive black hole (MBH) and periodically plows through its accretion disk.
In this work, we compute the gravitational wave (GW) signals emitted by such EMRI systems.
We find that the physical drag and perturbations due to shock caused by the orbiter-disk collisions leave a distinct imprint on the emitted waveforms.
Rather than the smooth, monochromatic evolution observed in vacuum systems, these interactions excite non-discrete modes that manifest as subtle shifts in the orbital frequency and as high-frequency ``tails'' in the signal spectrum. We demonstrate as an example outcome of our model that a specific QPE source RX J1301.9+2747 could be detectable by future space-based GW detectors, provided the orbiter maintains a moderate eccentricity of approximately $0.25$ and a mass exceeding $35\, M_\odot$. Our analysis shows that the signal-to-noise ratio for these events would be high enough to clearly distinguish them from standard vacuum EMRIs. Consequently, GW observations offer a powerful tool to probe the dense environments surrounding MBHs and could give further insight into the elusive origins of QPEs.
\end{abstract}


\section{Introduction} 
Quasiperiodic eruptions (QPEs) are a fascinating and relatively new phenomenon in high-energy astrophysics.
As of now, about ten QPE sources have been identified across various galactic nuclei hosting massive black holes (MBHs)~\citep{Arcodia2021, Miniutti2019, Giustini_2020, Chakraborty2021, Arcodia2021, Miniutti2022, Miniutti2023, Quintin_2023, Giustini_2024, Arcodia_2024, Chakraborty2025b, Arcodia_2025}, and they are characterized by luminous soft X-ray bursts ($\sim10^{42\text{--}43}\,\mathrm{erg\,s^{-1}}$) lasting $\sim1\,\mathrm{ks}$, with quasi-periodic recurrence times of $\sim10-100\,\mathrm{ks}$ and variable flare luminosities.
QPEs have rapidly risen to prominence in high-energy astrophysics since their discovery, with their physical origin remaining unresolved and actively investigated.
The proposed models include the transits of stellar-mass objects (SMOs) across an accretion disk, repeated partial tidal disruptions of an orbiting star, active-galactic-nuclei (AGN) disk instabilities, super-Eddington accretion disks with Lense-Thirring precession, mass transfer from an orbiting SMO to a MBH, gravitational lensing from close binary MBH systems, etc \citep{Karas_2001, Karas2004, Subr2004, Wevers_2017, Wevers_2019, king2020, Sukova_2021, Xian_2021, Ingram2021, King2022, Krolik2022, Metzger_2022, Wang_2022, Chen_2023, Lu2023, Middleton_2025, Chakraborty_2025}.
In this letter, we focus on the most well-studied ``orbiter disk interaction'' (ODI) model, in which the QPE flares are produced as a SMO closely orbits a MBH and continuously collides with the accretion disk of the MBH \citep{Zurek_1994, Dai2011, Franchini2023, Linial2023, Linial_2024, Chakraborty_2025}.
The ODI model is favored for its ability to naturally explain key observational features such as alternating flare amplitudes and quasi-periodic recurrence times.
This quasi-periodicity arises as the SMO on a quasi-Keplerian orbit around a MBH iteratively traverses a longer arc followed by a shorter arc.
This produces alternating time intervals between disk-crossing collisions that trigger flares, naturally creating the ``long-short-long-short'' recurrence pattern \citep{Linial_2024, Franchini2023}.
\begin{figure}[tbp]
    \includegraphics[width=0.47\textwidth]{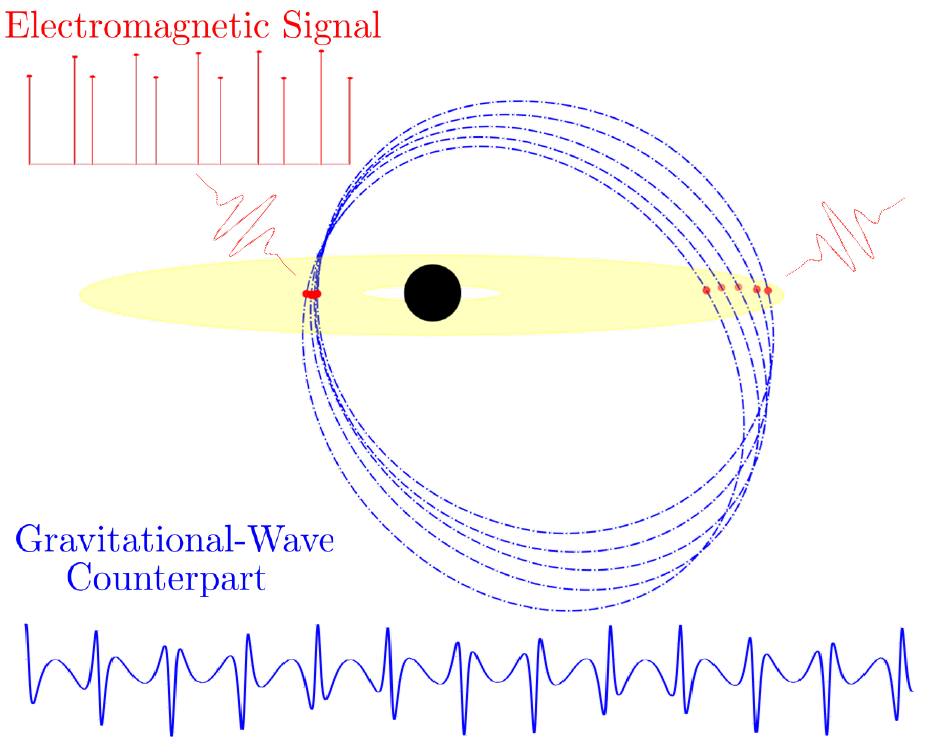}  
    \caption{Schematic depiction of an EMRI system where a SMO repeatedly collides with an accretion disk whilst orbiting a MBH.
These collisions produce EM flares modulated by relativistic precession and orbital inclination.
GWs emitted by the EMRI encode spacetime curvature perturbations, enabling multi-messenger probes of the orbital dynamics.}
    \label{QPE_GW_FIG}
\end{figure}

While numerous models have been proposed to explain the electromagnetic (EM) signature of QPEs, their multi-messenger implications remain largely underexplored.
\LL{Intriguingly, QPE systems described by the ODI model share key characteristics with extreme-mass-ratio inspirals (EMRIs), which are binaries where a SMO spirals towards a MBH emitting gravitational waves (GWs) \citep{Amaro_Seoane_2018}. However, it is crucial to note that definitive observational evidence confirming this QPE-EMRI connection is currently lacking, and some QPE sources exhibit characteristics that challenge a straightforward EMRI interpretation \citep{Chakraborty_2025}. In fact, this lack of conclusive evidence highlights the necessity of multi-messenger observations. Detecting the accompanying GWs from these systems would serve as a definitive test to verify the star-disk collision model and solidify the QPE-EMRI link.}
In this paper, we study these QPE/EMRI systems with ODIs model, focusing on eccentric orbits ($e > 0$).
Such systems are prime targets for space-based GW detectors like LISA, TianQin, and Taiji, operating in the frequency range $0.1\text{--}100\,\mathrm{mHz}$ \citep{LISA_2024, TianQin_2021, taiji_2015, Torres-Orjuela_2024, Torres-Orjuela_2024b}.
A joint EM-GW detection would establish QPEs as multi-messenger laboratories for probing accretion physics, strong-field gravity, and EMRI environments.
Figure.~\ref{QPE_GW_FIG} illustrates this framework, showing the interplay of EM signals from the ODI and the emission of GWs from the EMRI.
The ODIs extract energy and angular momentum from the EMRI, changing the orbital morphology and altering the EM and GW signals over observations.
Prior studies have provided estimates on the signal-to-noise ratio (SNR);
however, no current work consistently includes ODI effects when computing the GW signals.
\cite{Chen2022, Kejriwal_2024, Spieksma_2025, Zhan_2025, Suzuguchi_2025} found that by treating current QPEs as vacuum EMRIs on nearly circular orbits, the SNRs are far too low to be detectable by space-based GW detectors.
However, GW emission from eccentric binaries excites harmonics at higher frequencies, which can enhance detectability as they enter the sensitivity band of space-based detectors~\citep{Peters_1963, Peters_1964}.
While earlier work has suggested that high eccentricities might amplify GW signals from EMRIs, the applicability of this mechanism to QPEs remains unexplored~\citep{Amaro_Seoane2007}.
Moreover, we find that ODIs excite non-discrete GW modes, which give rise to high-frequency tails.
\LL{Although environmental effects typically reduce the match with vacuum-GR templates, which may worsen the detectability if purely vacuum templates are used, they can inherently enhance the intrinsic power of the GW signal at specific frequencies. Proper modeling of these environmental deviations is therefore necessary to recover and potentially enhance the detectability of such GWs~\citep{Torres-Orjuela_2021, Torres-Orjuela_2021b, Destounis_2021, De_Amicis_2024, Duque_2025, Dyson_2025}. However, these effects have never been explored in the context of QPEs.}
In this paper, we account for these ODI effects and gauge the detectability of the GWs emitted by observed QPE candidates.
Furthermore, we use the mismatch to determine whether the QPE GW signals can be differentiated from vacuum-EMRI GWs.
\section{Observational Constraint on Quasi-Periodic Eruptions}
\begin{figure}[tbp]
    \centering
    \includegraphics[width=0.49\textwidth]{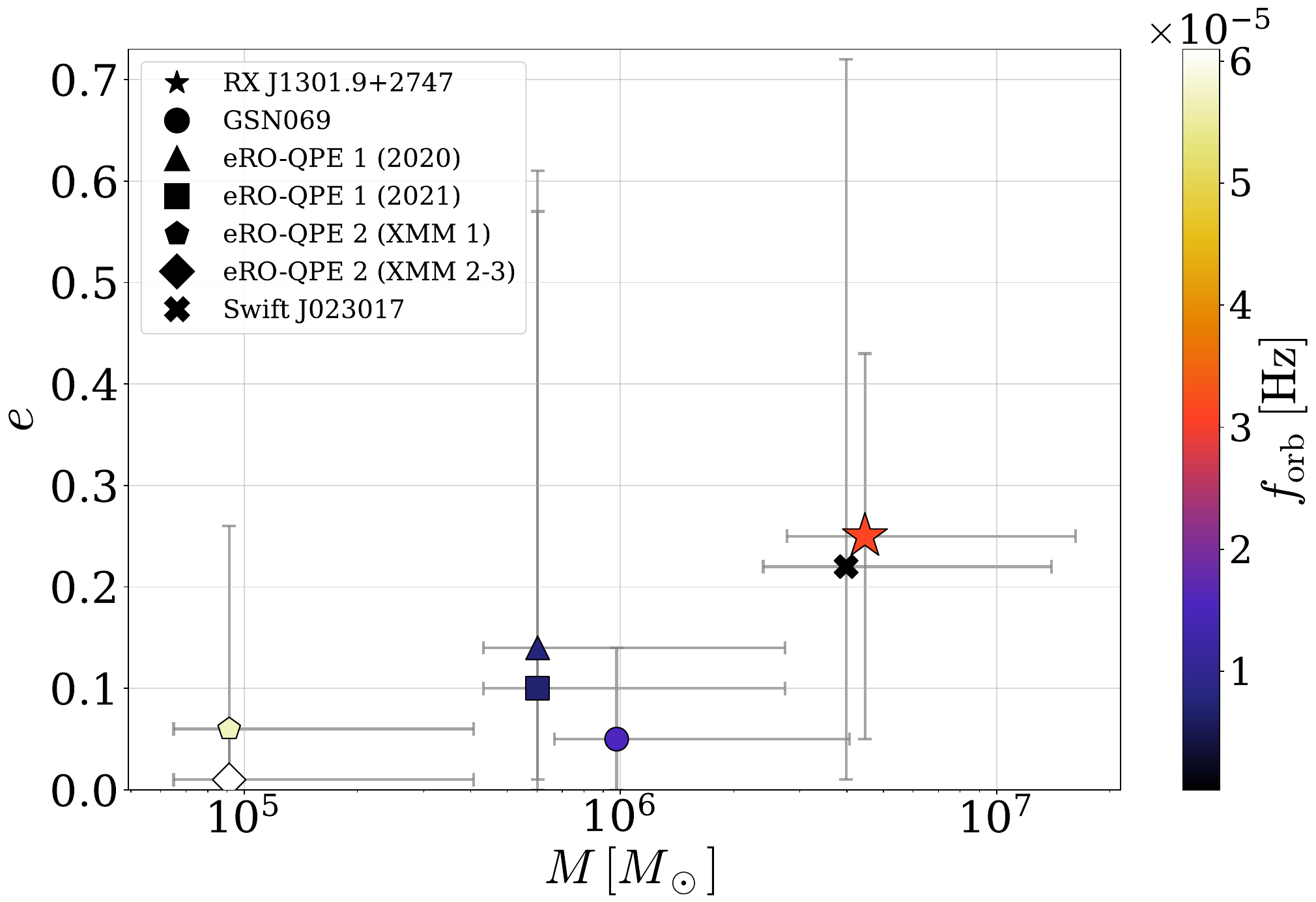}
    \caption{EMRI parameters of five QPEs with data obtained from \cite{Zhou_2024b}, where the $x$-axis is the MBH mass, and the $y$-axis is the orbital eccentricity.
The color encodes the orbital frequency of the QPE, and the error bars are the 2-$\sigma$ confidence interval.}
    \label{QPE_Constraints}
\end{figure}

The EM flares of the observed QPEs can be used to constrain the orbital information of the SMOs.
In the ODI model, the SMO collides twice with the disk per orbit, and hence the recurrence timescale of QPE flares corresponds to roughly half of the orbital period of the SMO, which in turn allows one to calculate the frequency and strength of the GW signals.
Another important parameter for the GW calculation is the eccentricity of the orbit, which can also be constrained using the pattern of the recurrence times observed in QPEs.
Generally speaking, a tighter orbit, a higher eccentricity, and a larger SMO mass should enhance the detectability of GWs emitted by EMRIs~\citep{Zhang_2023, Pitte_2023, Khavlvati_2025}.
Constraining QPE orbital parameters is complex due to degeneracies arising from relativistic precession.
In this letter, we directly adopt the parameter-fitting results from \cite{Zhou2024, Zhou_2024b} in which the authors fitted the QPE timings with a geodesic in Kerr spacetime.
The orbital parameters are fitted using the data from XMM Newton, Chandra, eROSITA, Swift-XRT~\citep{Dewangan_2000, Miniutti2019, Miniutti2022, Miniutti2023, Evans_2023, Guolo_2024, Pasham_2024, Chakraborty_2024, Chakraborty_2025}, and align well with existing QPE literature~\citep{Stone_2016, Kochanek_2016, Wevers_2017, Wevers_2019, Xian_2021, Zhou2024}.
Figure.~\ref{QPE_Constraints} shows the fitted orbital eccentricities $e$ and frequencies $f_{\mathrm{orb}}$ by \cite{Zhou2024, Zhou_2024b}, as well as the MBH masses $M$ obtained from the $M-\sigma$ relation \citep{Faber_Jackson1976, McConnell2011, Kormendy_2013} for five observed QPEs.

Analyses treating QPEs as EMRIs (while not accounting for ODIs) suggest that QPEs with very low eccentricities ($e\lesssim0.05$) do not produce GWs observable using space-based GW detectors \citep{Xian_2021, Pasham_2024}.
Indeed one sees in Figure.~\ref{QPE_Constraints} that all QPE orbital frequencies are in the range $0.01-0.1\mathrm {mHz}$, placing the dominant mode of GW emission for circular orbits ($\omega_{\mathrm{GW}} = 4\pi f_{\mathrm{orb}}$) below the sensitivity band of space-based detectors~\citep{LISA_2022, TianQin_2024}.
However, there exists a large uncertainty in the estimation of orbital parameters due to the limited number of cycles observed.
The literature suggests that at least some QPEs can have moderate orbital eccentricities ($e\gtrsim 0.2$)~\citep{Zhou2024, Zhou_2024b}, which possibly generate GWs with a significant fraction of their higher harmonics lying inside the detection band~\citep{Cornish_2003, Barack2007, Amaro_Seoane2007, Klein_2016, Feng_2019}.
In this letter, we focus on \LL{RX J1301.9+2747. We select this source not only because it has one of the largest reported orbital eccentricities ($e= 0.25^{+0.18}_{-0.20}$) and a large number of observed cycles ($N\sim10$), but also because it hosts a more massive MBH, resides in a closer host galaxy, and exhibits a higher orbital frequency compared to other candidates—all of which act synergistically to enhance its overall GW detectability. We caution, however, that RX J1301.9+2747 displays an erratic eruption pattern in both timing and luminosity that may be difficult to fully explain with a pure ODI model, as some stray eruptions elude simple EMRI-QPE fits \citep{Zhou2024}. Despite these caveats, its inferred parameters provide a highly compelling testbed for our GW detectability framework.}
There is a large uncertainty in the SMO mass, as it does not alter the QPE's periodicity due to the large mass ratio.
Therefore, in our results, we use the entire mass range of SMO, as LIGO has detected the black holes in the intermediate mass range ($\mu\gtrsim150M_{\odot}$)~\citep{Abbott_2020}.
\section{Methodology}
\subsection{Model for Orbiter-Disk Interactions}
We adopt the model by \cite{Zhou2024}, which calculates the dynamical friction produced by a SMO crossing an accretion disk as 
\begin{equation}\label{Chandrasekhar}
    \delta \E_{\mu} = 4\pi \ln|\Lambda|
\cdot \frac{G^2\mu}{v^2} \cdot \frac{\Sigma}{\sin I},
\end{equation}
where $\delta \E_{\mu}$ is the energy dissipated per unit mass, $\mu$ is the SMO mass, $v$ is the collision velocity, $\Sigma$ is the surface density of the disk, $I$ is the orbital inclination with respect to the equatorial plane, and $\Lambda = b_{\mathrm{max}}/b_{\mathrm{min}}$ is the ratio between the maximum and minimum impact parameter.
We use $b_{\mathrm{max}} = r_a$ and $b_{\mathrm{min}} = r_S$, where $r_a$ and $r_S$ are radial distances from the apocenter of the orbit and the separatrix in Kerr spacetime, respectively~\citep{Stein_2020}.
Assuming that the quiescent luminosity between QPE flares stems from disk emission, we use the $\alpha$-disk model to model the disk's surface mass density~\citep{Shakura1973}
\begin{equation}\label{SS}
    \Sigma = 5.2\alpha^{-\frac{4}{5}}\dot{M}_{16}^{\frac{7}{10}}M^{\frac{1}{4}}R_{10}^{-\frac{3}{4}}f^{\frac{14}{5}},
\end{equation}
where $\dot{M}_{16}:= \dot{M}/(10^{16}~\mathrm{g\,s}^{-1})$ with $\dot{M}$ denoting the accretion rate, $M$ is the MBH mass, $R_{10}:= R/(10^{10}~\mathrm{cm})$ with $R$ being the distance from the MBH, and $\alpha$ is the viscosity parameter for which we choose a fiducial value of $0.01$.
$f:= (1 - \sqrt{GM/(2\eta R)})^{1/4}$ arises from the inner boundary condition ensuring that the viscous stress vanishes at the inner edge of the disk, and $\eta$ is set to be 0.1, which is the standard accretion efficiency.
\subsection{Altered Kerr Geodesics}
For a BH of mass $M$ and spin $a$, the line element is given by the Kerr solution, and in Boyer-Linquist coordinate $(t,r,\theta,\phi)$, this is
\begin{equation}\label{Kerr_metric}
   \begin{gathered}
        \dd s^2=-\left(1-\frac{2Mr}{\Sigma}\right)\dd t^2-\frac{4Mar\sin^2\theta}{\Sigma}\dd t\dd\phi+\frac{\Sigma}{\Delta}\dd r^2\\
        +\Sigma\;\dd\theta^2+\left[(r^2+a^2)\sin^2\theta+\frac{2Mr}{\Sigma}a^2\sin^4\theta\right]\dd\phi^2,\\
   \end{gathered}
\end{equation}
where $\Sigma=r^2+a^2\cos^2\theta$, and $\Delta=r^2+a^2-2Mr$.
The geodesics for the metric in Eq.~\eqref{Kerr_metric} can be completely characterized by the constants of motion $(\E, \AM, \Q)$ which denote the specific energy, $z$-angular momentum, and Carter constant, respectively~\citep{Carter_1968}.
The large mass ratio between the MBH and the SMO allows us to evolve $(\E, \AM, \Q)$ associated with GW emission using perturbative methods.
The SMO is treated as a point mass which perturbs the Kerr metric via a gravitational self-force, and inspirals due to GW radiation.
This gravitational radiation and self-force can be computed using the Teukolsky equation~\citep{Teukolsky1972, Teukolsky1973, Teukolsky1974} (cf. Appendix~\ref{appendix_a}).
\LL{Throughout this letter, we use the \texttt{FastEMRIWaveforms} package~\citep{Barack_2004, Babak_2007, Chua_2015, Chua_2017, Chua_2019, Stein_2020, Fujita_2020, Chua_2021, Katz_2021, Speri_2024} to compute the GW signals and the adiabatic inspiral due to GW radiation~\citep{Schmidt_2002, Fujita_2020, Hughes_2024} between disk collisions.}
Whenever the SMO crosses the disk, we update the Kerr geodesics constants of motion $(\E, \AM,\Q)\to(\E', \AM',\Q')$~\citep{Carter_1968, Bardeen_1972} according to the ODI model
\begin{equation}\label{Kerr_geodesic_evolution}
    \E'=\E-\delta \E,\;\;
\AM'=\AM-\delta \AM,\;\;
    \Q'=\Q-\delta\Q.
\end{equation}

Here $\delta \E=\delta \E_\mu$ is given by Eq.~(\ref{Chandrasekhar}) while $\delta \AM$ and $\delta\Q$ can be obtained by solving~\citep{Schmidt_2002, Flanagan_2007, Fujita_2020,  Hughes_2024}
\begin{equation}\label{Jacobian}
\begin{pmatrix}
    \delta p\\
    \delta e \\
    \delta x_I\\
\end{pmatrix}
=\begin{pmatrix}
    \frac{\partial p}{\partial \E}&\frac{\partial p}{\partial \AM}&\frac{\partial p}{\partial \Q}\\
    \frac{\partial e}{\partial \E}&\frac{\partial e}{\partial \AM}&\frac{\partial e}{\partial \Q}\\
    \frac{\partial x_I}{\partial \E}&\frac{\partial x_I}{\partial \AM}&\frac{\partial x_I}{\partial \Q}\\
\end{pmatrix}\begin{pmatrix}
    \delta \E\\
    \delta \AM \\
    \delta \Q\\
\end{pmatrix},
\end{equation}
which is the mapping of the Kerr geodesic constants $(\delta \E,\delta  \AM, \delta 
\Q)$ to the orbital parameters of quasi-Keplerian orbits $(\delta p ,\delta e, \delta x_I)$ ($p$: the semi-latus rectum, $e$: the orbital eccentricity, $x_I:=\cos I$: the $\cos$ of the inclination $I$) using the Jacobian matrix (cf. Appendix~\ref{appendix_b}). The system of equations in Eq.~\eqref{Jacobian} is underdetermined, so two further constraints are imposed for this calculation.
First, since the ODI is modeled as a drag, we assume this drag exerts forces parallel to the orbital motion, damping the SMO's velocity while not changing its direction, i.e., $I$ should remain constant or $\delta x_I=0$.
This approximation is valid in the thin-disk regime, when the SMO is immersed in the disk for a very short time, and the supersonic regime, where the disk's velocity is negligible compared to the SMO's orbital velocity~\citep{Duque_2025}.
Second, we use Kepler's equation $r=p/(1+e\cos\chi)$ and set $\delta r=0$, to prevent unphysical effects where the SMO abruptly changes its radial position after the ODI.
This leads to $\delta p = p\cdot\delta e\cos\chi/(1+\cos\chi)$. Furthermore, since the ODI occurs in the equatorial plane and given that Eq.~\eqref{Jacobian} is $\phi$-independent (due to the axisymmetry of the Kerr metric), $\delta\theta=\delta\phi=0$.
These calculations are done iteratively over LISA's planned 4-year operation time.
\section{Results}
\subsection{Imprints on Gravitational Waves from Disk Collisions}
\begin{figure}[htbp]
    \includegraphics[width=0.4834\textwidth]{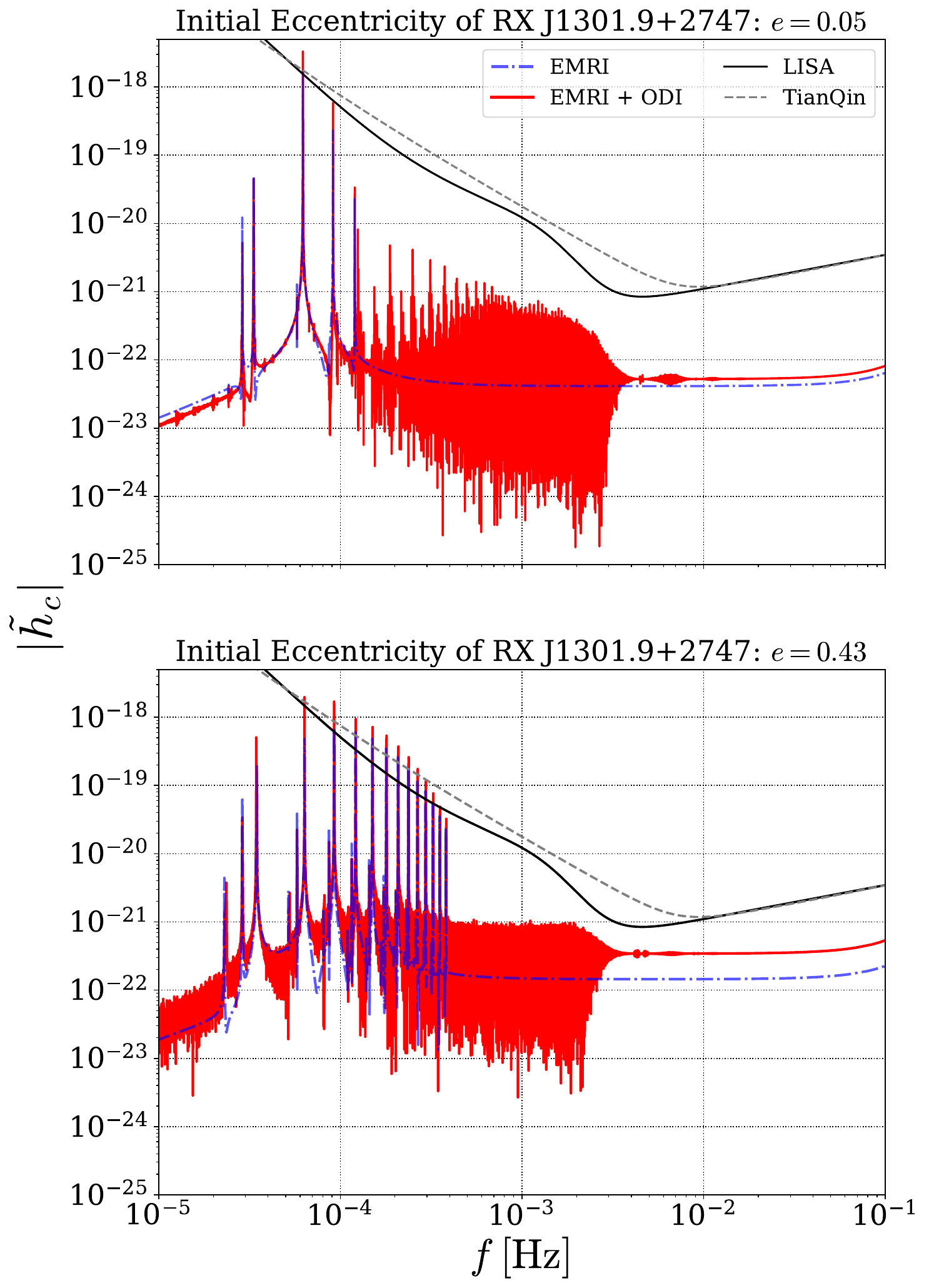}
    \caption{The characteristic strain over LISA's operation time of four years for a vacuum EMRI (dashed blue line) versus an EMRI with ODIs (solid red line) using the smallest (top panel) and largest (bottom panel) inferred initial eccentricity $e_0=0.25^{+0.18}_{-0.20}$ of RX J1301.9+2747 with the noise curves of LISA (black line) and TianQin (dashed gray line).
The initial parameters are $M=4.47\times 10^6M_{\odot}$, $\mu = 30 M_{\odot}$, $a=0.61M$, $p_0=55.5 M(1-e_0^2)$, $x_{0}=\cos I_0=0.63$, and the luminosity distance $d_L=100\,{\rm Mpc}$.
}
\label{QPE_GW_Comparison}
\end{figure}

GW radiation drives inspiral and circularization for EMRIs~\citep{Peters_1963, Peters_1964}.
The energy loss due to GW radiation causes the orbit to become more tightly bound, reducing the semi-major axis.
The angular momentum loss damps eccentricity because GWs are emitted more strongly during close approaches in eccentric orbits.
ODIs accelerate the inspiral, as shocks introduce additional orbital energy loss.
While the energy dissipated in a single ODI is small compared to the orbital binding energy, over the course of tens of thousands of cycles, the total energy and angular momentum dissipated due to ODIs can accumulate and leave observable signatures on the GW signals.
ODIs also accelerate circularization. For a Keplerian orbit, we have $p=\AM^2/M$ and $e^2=1+2\E\AM^2/M^2$.
The orbital eccentricity varies as $\delta e=(p/M\cdot\delta \E-(1-e^2)/\sqrt{pM}\cdot\delta \AM)/e$. The first term dominates because $p$ amplifies ODI energy perturbations while suppressing angular momentum effects, as $\delta \E$ and $\delta \AM$ are of similar order of magnitude.
Therefore, ODIs reduce $e$ by introducing a negative $\delta \E$ effect which accelerates the orbital circularization.
We plot in Figure.~\ref{QPE_GW_Comparison} the characteristic strain for RX J1301.9+2747 using the minimal and maximal eccentricities constrained using QPE signals ($e_0=0.05,0.43$, see Figure.~\ref{QPE_Constraints}), considering a vacuum EMRI and an EMRI with ODIs.
One can see that the more eccentric orbit with $e_0=0.43$ can place the higher harmonics into LISA's detectability band, and the ODIs amplify the power of the spectrum at these discrete higher harmonics.
Moreover, the spectra for EMRIs with ODIs have high-frequency tails of non-discrete modes.
This improves the GW detectability by enhancing the power of the spectrum at higher frequencies, as the characteristic strain for EMRIs with ODIs is higher than that of vacuum EMRIs.
To understand the GW waveform of EMRIs with ODIs, we consider a simple case where an EMRI abruptly changes its orbital frequency, semi-latus rectum, and eccentricity.
This alters the GW frequencies $\omega_{nmk}$ and amplitude $\mathcal{A}_{\ell nmk}$, where $\ell$ is the orbital angular momentum mode index, and $n$, $m$, and $k$, are the radial, azimuthal, and polar mode indices, respectively.
For vacuum EMRIs, the strain can be written as a sum of multiple modes with amplitudes $\mathcal{A}_{\ell nmk}$ and phases $\Phi_{nmk}$
\begin{equation}
     h(t) \propto \sum_{\ell,m,n,k}\mathcal{A}_{\ell nmk}(t)e^{-i\Phi_{nmk}(t)}.
\end{equation}
If there is a sudden change in the frequency $\omega_{nmk} := \mathrm{d}\Phi_{mnk}/\mathrm{d}t$ from $\omega_{nmk}$ to $\omega'_{nmk}=\omega_{nmk}+\delta\omega_{nmk}$ and a sudden change in amplitude $\mathcal{A}_{\ell nmk}'=\mathcal{A}_{\ell nmk}+\delta \mathcal{A}_{\ell nmk}$ at $t=t_c$, the integral for the Fourier transform needs to be split into two parts, one $-\infty<t<t_c$ and another $t_c<t<\infty$.
Using the stationary phase approximation and the Sokhotski-Plemelj theorem, the result is a frequency spectrum $\tilde{h}(\omega)\propto\sum_{\ell,m,n,k}\tilde{h}_{\ell nmk}$ split into three parts
\begin{equation}\label{GW_spectra}
\begin{split}
        \tilde{h}_{\ell nmk}=&\;\tilde{\mathcal{A}}_{\ell nmk}(\omega)\delta(\omega-\omega_{nmk}) \\
        &+\tilde{\mathcal{A}}_{\ell nmk}'(\omega)\delta(\omega-\omega_{nmk}')\\
    &+\frac{e^{-i\omega_{nmk}t}}{i\omega}\left(\delta\tilde{\mathcal{A}}_{\ell nmk}(\omega)+\frac{\delta\omega}{\omega}\tilde{\mathcal{A}}_{\ell nmk}(\omega)\right).\\
\end{split}
\end{equation}
The first and second terms in Eq.~\eqref{GW_spectra} correspond to the GW fundamental frequencies and their mode shifts due to the altered dynamics, respectively, where the tildes denote the Fourier transform.
From Figure.~\ref{QPE_GW_Comparison}, one sees that the lower harmonics of EMRIs with ODIs align with those of vacuum EMRIs without a clear splitting of the GW modes.
This is because the shifts in the fundamental frequencies $\omega_{nmk}$ are small, as $\delta\omega_{nmk}\propto p^{-5/2}\delta p$ and the shift in semi-latus rectum is small $\delta p/p\sim 10^{-5}$ for EMRIs far from the merger.
The non-discrete (`noisy') part of the spectrum for EMRIs with ODIs corresponds to the last term in Eq.~\eqref{GW_spectra}.
It stems from discrete reductions in $(\E, \AM, \Q)$ during ODIs (cf. Eq.~\eqref{Kerr_geodesic_evolution}), which cause abrupt declines in $p$ and $e$, and give rise to high-frequency tails commonly seen in EMRI glitches and the ringdown phase~\citep{Ching_1995, Han_2019, Destounis_2021, De_Amicis_2024, Islam_2024}.
This effect is analogous to striking a bell, where the impulse excites vibrations at multiple frequencies, producing a transient ``ring''  that decays over time.
This interpretation is consistent with our model, as the impulsive deceleration of the SMO during an ODI excites transient oscillations in the GW signal, differing from the scenario of vacuum EMRIs in which $p$ and $e$ evolve smoothly under gravitational radiation.
\subsection{Detectability of Gravitational Waves from Quasi-Periodic Eruptions}
\begin{figure}[htbp!]
\centering
    \includegraphics[width=0.483\textwidth]{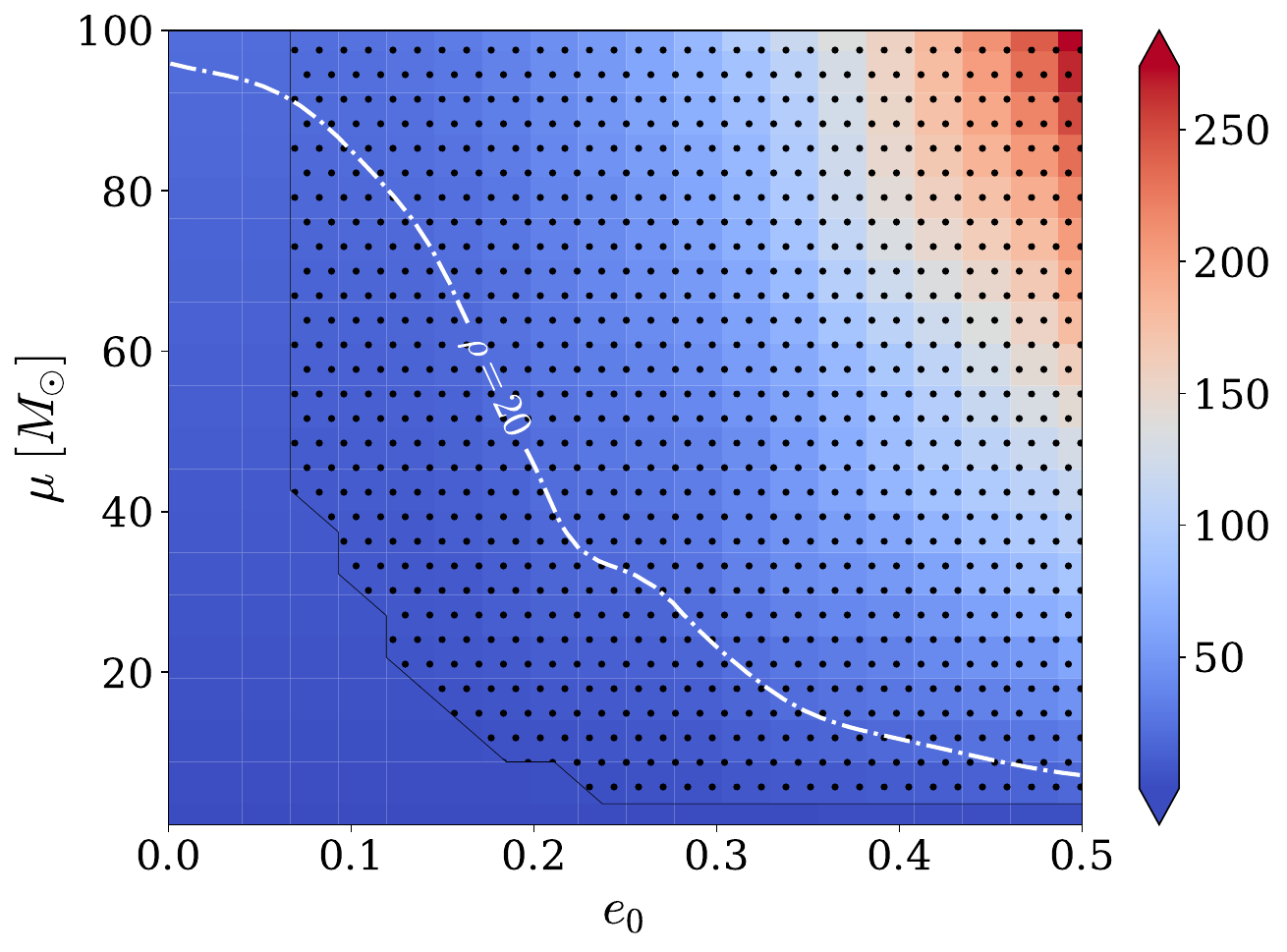}
    \includegraphics[width=0.483\textwidth]{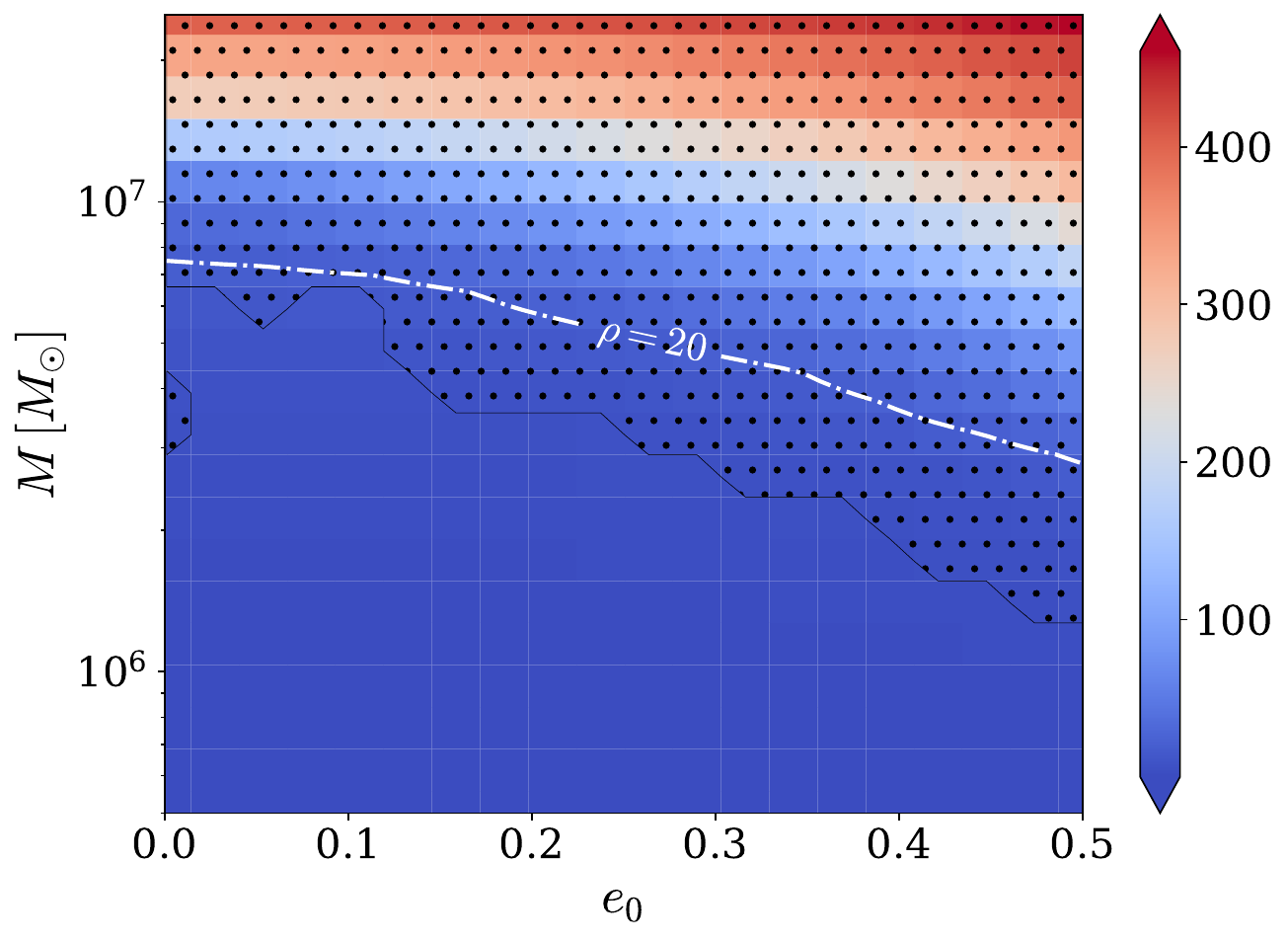}
    \includegraphics[width=0.483\textwidth]{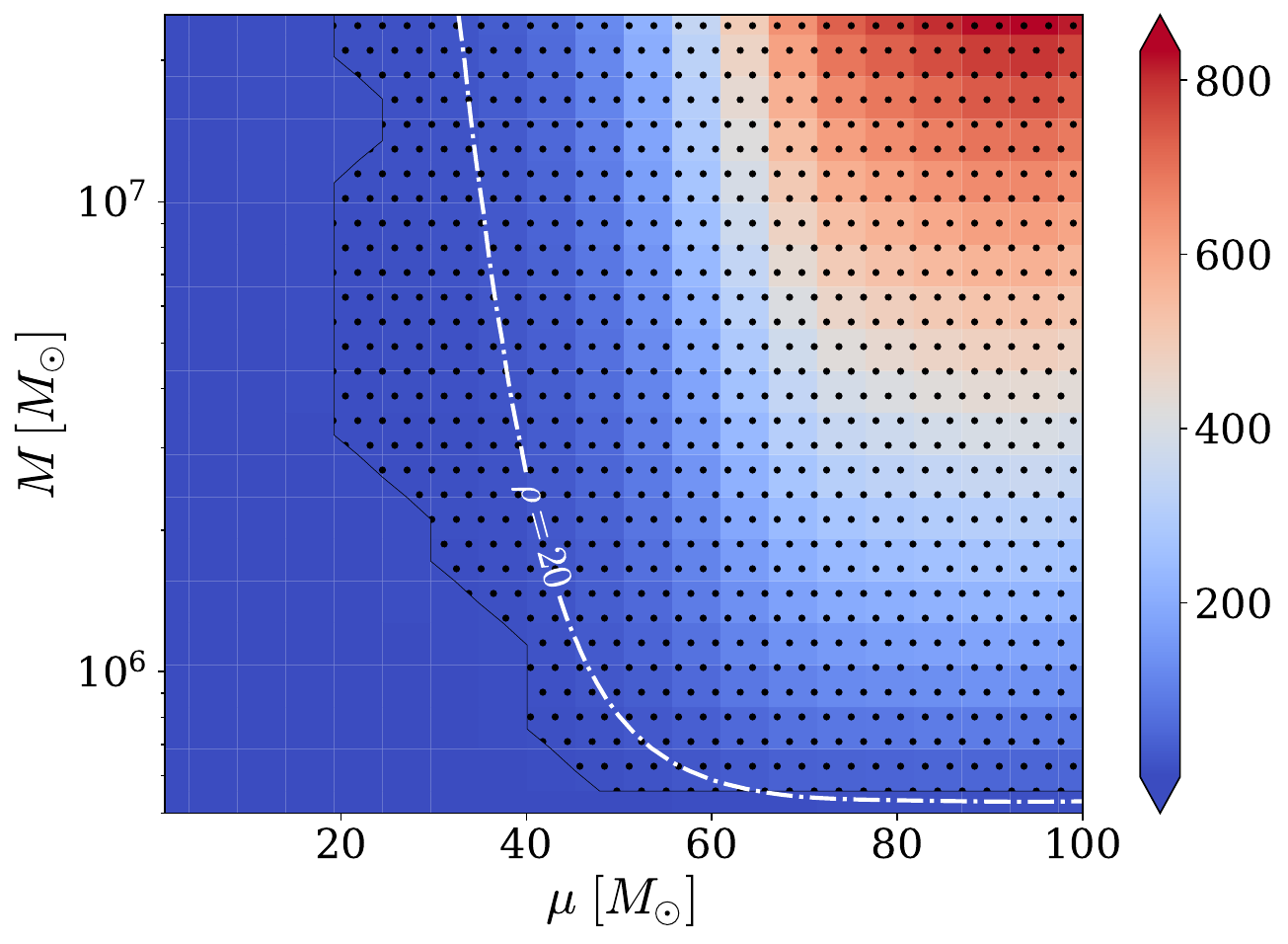}
\caption{SNR of an EMRI with ODIs for varying orbiter mass $\mu$ and initial eccentricity $e_{0}$ (top), MBH mass $M$ and $e_{0}$ (middle), and $M$ and $\mu$ (bottom).
Fiducial values ($M = 4.47 \times 10^{6} M_{\odot}$, $e_0 = 0.25$) are explicitly based on QPE RX J1301.9+2747.
MBH spin $a = 0.61M$, luminosity distance $d_{L} = 100$~Mpc, and initial orbital inclination cosine $x_{0} = 0.63$ are fixed.
The white dashed-dotted line marks the SNR value $\rho = 20$;
the hatched region indicates where $\rho$ allows differentiation from a vacuum EMRI (cf. Eq.~\eqref{SNR_Mismatch}).}
\label{SNR_Mismatch}
\end{figure}
We use LISA's power spectral density (PSD)~\citep{Cornish_2003, Klein_2016, Feng_2019, Torres-Orjuela_2024} and the luminosity distance $d_L=100\,{\rm Mpc}$ from observations~\citep{Giustini_2024} to compute the SNR of an EMRI with ODIs and its match with a vacuum EMRI with otherwise equal parameters.
\LL{Since our time-domain EMRI waveforms are continuously tracked over an extremely long duration (spanning the continuous four-year observation window of LISA), the spectral leakage arising from finite-boundary effects is negligible. Thus, the enhanced power observed at high frequencies is a genuine physical imprint of the ODIs, rather than a numerical artifact.} The detectability of a GW signal $h$ is quantified by its optimal SNR~\citep{Finn_1992, Feng_2019}
\begin{equation}
    \rho = \sqrt{\langle\tilde{h}|\tilde{h}\rangle},
\end{equation}
where $\langle\tilde{h}|\tilde{\bar{h}}\rangle := 4\mathfrak{Re}\left\{\int^{f_{\mathrm{high}}}_{f_\mathrm{low}}\mathrm{d}f\tilde{h}^*(f)\tilde{\bar{h}}(f)/{S_n(f)}\right\}$ is the noise-weighted inner product of the Fourier transformed waveforms $\tilde{h}$ and $\tilde{\bar{h}}$, $f_{\mathrm{low}}\leq f \leq f_{\mathrm{high}}$ is the frequency range of the detector (we use $f_{\mathrm{low}}=0.1\;\mathrm{mHz}$ and $f_{\mathrm{high}}=100\;\mathrm{mHz}$), and $S_n$ is the PSD of the noise.
An EMRI is usually considered to be detectable by space-based detectors if $\rho\geq20$~\citep{babak_gair_2017}.
The similarity between the waveforms of the ODI-perturbed EMRIs $h$ and vacuum EMRIs $\bar{h}$ is quantified by their match $\mathcal{M}$~\citep{Owen_1996, Mohanty_1998, Purrer_2020, Pizzati_2022} 
\begin{equation}
    \mathcal{M}(h,\bar{h})=\frac{1}{\sqrt{\rho\bar{\rho}}}\langle\tilde{h}|\tilde{\bar{h}}\rangle,
\end{equation}
and we can determine whether the two GW signals are distinguishable by the ``rule-of-thumb'' criteria ~\citep{Cutler_2007}
\begin{equation}\label{ROT}
    \rho > \sqrt{\frac{D}{2(1-\mathcal{M})}},
\end{equation}
where $D\approx 15$ is the number of source parameters.
In Figure.~\ref{SNR_Mismatch}, we show the SNR of RX J1301.9+2747 for varying MBH mass $M$, SMO mass $\mu$, and initial eccentricity $e_0$.
We adopt the $M$ and $e_0$ inferred in \cite{Zhou_2024b}, and consider $\mu$ in the entire range of SMOs ($1\text{-}100\,M_\odot$), as $\mu$ is not well constrained from QPE observations.
We use the current inferred orbital parameters from \citet{Zhou2024, Zhou_2024b}, instead of the projected values for mid-2030s detectors, as these parameters deviate slowly over 10 years (cf. Appendix~\ref{appendix_b}).
Also, our extended parameter space in Figure.~\ref{SNR_Mismatch} already accounts for these projected orbital parameters.
The SNR increases with $M$, $\mu$, and $e_0$, as $|\tilde{h}_c|\propto\mu M$, and larger values for $e_0$ give rise to higher harmonics which are pushed into the sensitivity band of LISA.
From the bottom panel of Figure.~\ref{SNR_Mismatch}, we see that RX J1301.9+2747 will be detectable through GWs if $\mu\gtrsim35\,M_\odot$ for almost all $M$ considered if $e_0=0.25$, while $\rho$ can go over 800 for $M\gtrsim10^7\,M_\odot$ and $\mu\gtrsim80\,M_\odot$.
However, even with lower SMO masses of $\mu\approx10\text{-}20\,M_\odot$, the QPE GW could still be detected if $e_0\gtrsim0.3$ and $M=4.47\times10^6\,M_\odot$ (cf. top panel of Figure.~\ref{SNR_Mismatch}).
Finally, we see from the middle panel of Figure.~\ref{SNR_Mismatch}, that for $\mu=30\,M_\odot$ EMRIs with $M\gtrsim3\times10^6\,M_\odot$ will be detected if $e_0\approx0.5$.
For almost all cases where a QPE emits detectable GWs, the SNR is also high enough to be differentiated from vacuum EMRIs.
Conducting the same calculation using TianQin's noise curve yields sub-detection SNRs, as TianQin is less sensitive to sources that emit low-frequency GWs (see Figure.~\ref{QPE_GW_Comparison}), such as high-mass MBHs.
However, TianQin could still be useful for detecting QPE sources that host low-mass MBHs~\citep{TianQin_2021, TianQin_2024}.
\section{Discussion and Conclusion}

In this letter, we modeled QPE bursts as EMRIs perturbed by ODIs to compute their GW signatures.
Our main findings are summarized as follows:
\begin{itemize}
    \item \textbf{Altered GW Signatures:} Physical drag and shocks from disk encounters significantly alter orbital dynamics and GW waveforms, enabling differentiation from vacuum EMRIs.
\item \textbf{Non-Dicrete Modes:} ODIs excite non-discrete GW modes appearing as high-frequency tails (Figure~\ref{QPE_GW_Comparison}, \LL{Figure~\ref{QPE_GW_FIG}}), enhancing characteristic strain and detectability.
\item \textbf{Detection Prospects:} RX J1301.9+2747 is a prime LISA candidate.
With orbiter mass $\mu \ge 35 M_{\odot}$ and eccentricity $e_{0} \approx 0.25$, its SNR allows detection and differentiation from vacuum EMRIs (Figure~\ref{SNR_Mismatch}).
\end{itemize}

Simultaneous detection of EM and GW signals from QPEs will play a crucial role in elucidating their physical origins.
While EM observations provide excellent constraints on the redshift and orbital period, GWs can substantially reduce the degeneracies caused by relativistic precession, providing tighter constraints on the SMO mass, MBH spin, and eccentricity.
Although a full Bayesian parameter estimation would provide a more robust assessment, the high predicted SNRs ($\rho \ge 20$) justify our use of a simple mismatch criterion to establish that ODI waveform distortions are discernible from vacuum GW signals~\citep{Cutler_2007}.
We recognize that current observational constraints carry large uncertainties, particularly regarding the MBH mass inferred from stellar dispersion velocities~\citep{Zhou_2024b, Zhou_2025, Zhou_2025b}.
However, our expanded parameter space accounts for these variances, and also agrees with more refined QPE-timing constraints~\citep{Chakraborty_2025}.

\LL{A limitation of our current methodology is the treatment of ODIs as instantaneous changes to the orbital parameters. While this introduces strict mathematical discontinuities in the waveform's amplitude and phase (Eq.~\eqref{GW_spectra})—manifesting as $1/\omega$ and $1/\omega^2$ spectral terms that extend to infinite frequencies—we justify this approximation physically. The accretion disk is assumed to be geometrically thin, and the orbiter crosses it at highly supersonic velocities. Specifically, the transit duration is roughly $\Delta t \approx H / (v \sin I)$, where $H$ is the disk thickness~\citep{Spieksma_2025}. For a thin disk with an aspect ratio $H/r \sim 10^{-2}$, the transit time is a minute fraction of the orbital period ($\Delta t/P \sim 10^{-3}$). In the frequency domain, averaging the drag over this finite thickness acts as a low-pass filter, which would smoothly suppress spectral power at extreme frequencies $f \gtrsim f_{\mathrm{cut}} \approx 1/\Delta t$. For the parameters of RX J1301.9+2747, $\Delta t \sim \mathcal{O}(10)\,\mathrm{s}$, placing the physical cutoff frequency $f_{\mathrm{cut}} \gtrsim 100\,\mathrm{mHz}$. Because this cutoff lies at the very upper boundary of the LISA sensitivity band ($0.1\text{--}100\,\mathrm{mHz}$), the smoothing effects are negligible within our observable window. Consequently, the rapid deceleration genuinely mimics a physical shock within the detector band, and the prominent high-frequency tails we observe are a robust physical consequence of this interaction rather than a pure artifact of the instantaneous-jump model.}

\LL{Furthermore, EMRIs within the LISA band are expected to occasionally pass through transient orbital resonances \citep{Isoyama_2022}, which can induce near-instantaneous jumps in orbital parameters and potentially mimic the high-frequency tail signatures of ODIs. However, this degeneracy actually works in our favor: purely gravitational orbital resonances do not produce electromagnetic signals. If QPEs arise from the ODI model, the coincident detection of periodic X-ray flares will cleanly break this degeneracy, allowing us to easily distinguish an ODI-perturbed EMRI from a vacuum EMRI undergoing a transient resonance.}

Future waveform modeling could improve upon this framework by incorporating near-identity transformations to accelerate inspiral computations~\citep{Van_de_Meent_2018, Lynch_2021, Lynch_2023, Drummond_2023, Lynch_2024}, as well as including fully-relativistic environmental effects that become significant in the late inspiral phases~\citep{Vicente_2025}.
Finally, while the EMRI-ODI model remains the most well-studied explanation yielding promising GW counterparts, other proposed mechanisms---such as partially stripped white dwarfs~\citep{Chen_2022} or precessing massive disks~\citep{Wessel_2021, Tsokaros_2022}---could also produce detectable signals.
Future studies should also explore whether the dense environments surrounding QPEs can further modify GW signals via backreactions~\citep{Dyson_2025, Polcar_2025}, an effect currently considered negligible for thin disks but potentially relevant in other accretion models.
\section{Acknowledgment}
We thank Xian Chen, Conor Dyson, Pau Amaro-Seone, Zhen Pan, Yi-Ming Hu, Tom Kwan, Samson Leong, Lars Lund Thomsen, and Zijian Zhang for the discussions.
We acknowledge support from the National Natural Science Foundation of China and the Hong Kong Research Grants Council (N\_HKU782/23, 17305523).
ATO acknowledges support by the National Science Foundation of China (No. W2533010).
ATO and LL were supported by Beijing Natural Science Foundation (No. IS25014).
RKC would like to acknowledge the financial support provided by the Anusandhan National Research Foundation (ANRF), a statutory body of the Department of Science and Technology (DST), Government of India, through the National Post-Doctoral Fellowship (NPDF) [Grant No. PDF/2025/004682]

\setcounter{equation}{0}

\appendix
\section{Orbital Evolution via Gravitational Radiation}\label{appendix_a}
As mentioned in the main text, the QPEs in our model are treated as EMRIs, where the orbital dynamics are governed by the gravitational radiation and discrete environmental drag.
The orbital evolution corresponding to GW radiation and its associated GW signals are computed using the Teukolsky master equation.
This equation describes linear perturbations of the Kerr metric by governing the evolution of the Weyl scalars \citep{Teukolsky1972, Teukolsky1973}.
The large mass ratio of EMRIs allows the SMO to be approximated point mass $\mu$ that perturbs the background curvature through a gravitational self-force~\citep{Teukolsky1973}.
\begin{figure*}[htbp!]
    \centering
    \includegraphics[width=\linewidth]{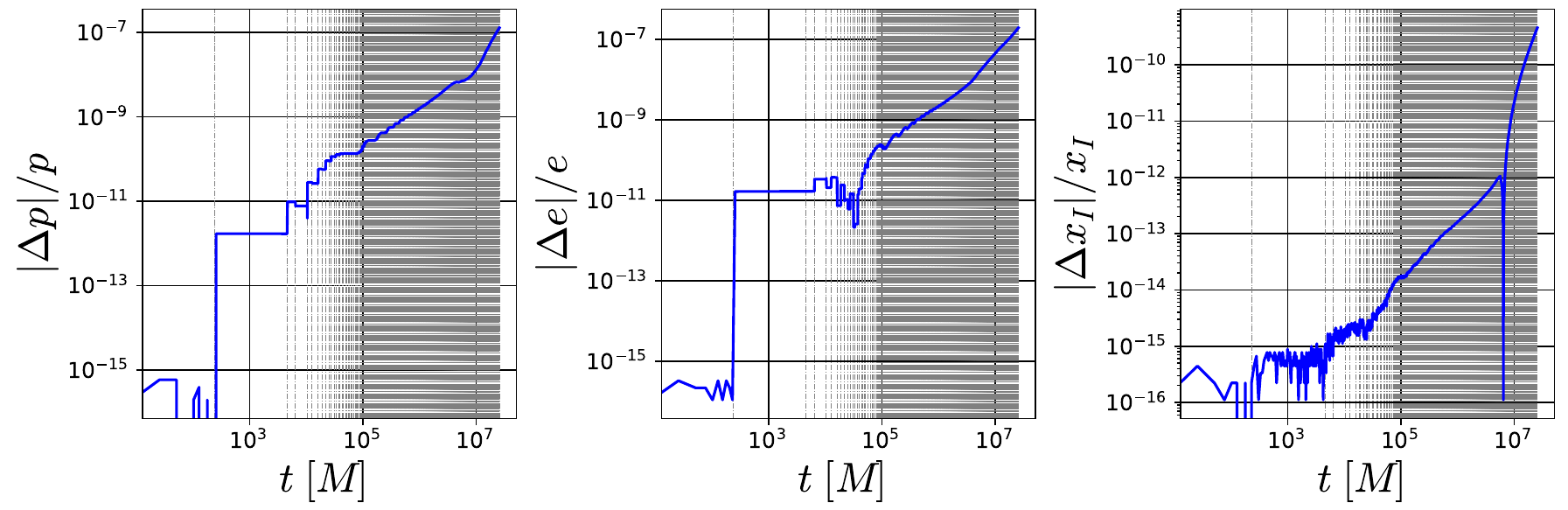}
    \caption{The relative difference in orbital parameters between a QPE EMRI perturbed by ODIs and a vacuum EMRI $(|\Delta p|/p, |\Delta e|/e, |\Delta x_I|/x_I)$.
The initial conditions for the orbit are $p_0=73.5 M(1-e_0^2)$, $e_0=0.25$, $x_{I0}=0.5$, $\Phi_{r0}=\Phi_{\theta0}=\Phi_{\phi0}=0$, and the BH masses and spins are $M= 10^{6}M_{\odot}$, $\mu=30M_{\odot}$, and $a=0.61M$.
The gray dash-dotted lines indicate the times when ODIs occur.}
    \label{QPE_orbit}
\end{figure*}
The Teukolsky equation in Boyer-Lindquist coordinates $(t,r,\theta,\phi)$, is written as follows~\citep{Teukolsky1972, Teukolsky1973, Teukolsky1974}

\begin{widetext}
    \begin{equation}\label{Teukolsky}
        \begin{gathered}
            \left[\frac{(r^2+a^2)^2}{\Delta}-a^2\sin^2\theta\right]\partial_t^2\psi+\frac{4Mar}{\Delta}\partial_{\phi}\partial_t\psi-\Delta^{-s}\partial_r\left(\Delta^{s+1}\partial_r\psi\right)+\left[\frac{a^2}{\Delta}-\frac{1}{\sin^2\theta}\right]\partial^2_{\phi}\psi
           -\frac{1}{\sin\theta}\partial_{\theta}\!\left[(\sin\theta)\partial_{\theta}\right]\\+2s\left[\frac{a(r-M)}{\Delta}+i\frac{\cos\theta}{\sin^2\theta}\right]\partial_{\phi}\psi+2s\left[\frac{M(r^2-a^2)}{\Delta}-r-ia\cos\theta\right]\partial_t\psi+(s^2\cot^2\theta+s)\psi=4\pi \mathfrak{T}(t,r,\theta,\phi),\\
        \end{gathered}
\end{equation}
\end{widetext}

where $s$ is the spin weight ($s=-2$ for outgoing GWs), $\psi=(r-ia\cos\theta)^4C_{\alpha\beta\gamma\delta}\;n^{\alpha}\bar{m}^{\beta}n^{\gamma}\bar{m}^{\delta}$ is the rescaled Weyl scalar constructed from the Weyl tensor $C_{\alpha\beta\gamma\delta}$ and the Kinnersly null tetrads $(l^{\alpha},n^{\alpha},m^{\alpha},\bar{m}^{\alpha})$~\citep{Newman_1961, Geroch_1975, Held_1975}, and 
$\mathfrak{T}$ is the source term constructed by acting the Newman-Penrose coefficients and their covariant derivatives, on the energy-momentum tensor $\mathcal{T}_{\alpha\beta}$ projected on the Kinnersly null tetrads~\citep{Teukolsky1972, Teukolsky1973, Teukolsky1974, Geroch_1975, Held_1975}.
The separability of the Teukolsky equation is a remarkable property stemming from the algebraic structure of the Kerr metric, which belongs to the class of vacuum Petrov type-D spacetimes~\citep{Teukolsky1973, Geroch_1975, Held_1975}.
This symmetry gives rise to an additional conserved quantity, namely the Carter constant $\Q$, which allows us to reduce the 4-dimensional partial differential equation into a set of uncoupled ordinary differential equations by decomposing $\psi$ into discrete frequency components and azimuthal harmonics~\citep{Teukolsky1972, Teukolsky1973, Drasco_2004, Drasco_2006} $(\ell m n k)$ 
\begin{equation}\label{separation_of_variables}
    \psi=\sum_{\ell, m,n,k}\mathcal{R}_{\ell mkn}(r)S_{\ell mkn}(\theta,\phi)e^{-i\omega_{mkn} t},
\end{equation}
where $\ell$, $m$, $n$, and $k$, are the orbital angular momentum, azimuthal, polar, and radial modes, respectively, and $\omega_{mnk} = m\Omega_\phi + n\Omega_r + k\Omega_\theta$ is the mode frequency corresponding to the periodicities of a stable bound orbit in Kerr spacetime.
The angular part can be decomposed into spin-weighted spheroidal harmonics $S_{lmnk}(\theta)$ which satisfies\footnote{Here, the mode indices have been omitted to save writing.}
\begin{equation}
\begin{split}
     &\left[ \frac{1}{\sin\theta} \partial_\theta (\sin\theta \partial_\theta) + a^2\omega^2\cos^2\theta \right.\\
     &\left.- \frac{(m-2\cos\theta)^2}{\sin^2\theta} + 4a\omega\cos\theta - 2 + A \right] S= 0,\\
\end{split}
\end{equation}
where $A$ is a separation constant.
The radial propagation is determined by the radial Teukolsky equation
\begin{equation}\label{Teukolsky_R}
    \Delta^{-s}\partial_r\left(\Delta^{s+1}\partial_r\mathcal{R} \right)+V\mathcal{R} =\mathcal{T},
\end{equation}
where 
\begin{equation}
    V(r)=-\frac{K^2-2is(r-M)K}{\Delta}+4is\omega r-\lambda
\end{equation}
is the Teukolsky potential, $K=\omega(r^2+a^2)-am$, $\omega=\omega_{mkn}$
$K = (r^2+a^2)\omega - ma$ and $\lambda = A + a^2\omega^2 - 2am\omega - 2$.
Eq.~\eqref{Teukolsky_R} can be solved by constructing a Green's function and integrating it over the source term~\citep{Hartle_1974, SN_1982, SN_1982b, Chandrasekhar_1975, Poisson_1993, Cutler_1993, Apostolatos_1993, Poisson_1993b, Poisson_1995, MST_1996, MST_1997, Campanelli_1997, Hughes_2000, Lo_2024}.
We model an EMRI with ODIs as a point-particle in Kerr spacetime that is perturbed by an ODI at $(t_i,\mathbf{x}_i)$.
Therefore, $\mathcal{T}^{\alpha\beta}=\sum_{i}\mathcal{T}^{\alpha\beta}_i$ with
\begin{equation}
    \mathcal{T}_i^{\alpha\beta} = \frac{\mu}{\sqrt{-g}}u_i^{\alpha} u_i^{\beta}\dot{\tau}_i \delta^3\left(\mathbf{x}(t) - \mathbf{x}_i\right) \Theta(t - t_i) \Theta(t_{i+1} - t),
\end{equation}
where $u^{\alpha}$ is the 4-velocity of the point mass $\mu$, $\delta^3(\cdot)$ is the 3-dimensional $\delta$-Dirac distribution, $\Theta(\cdot)$ is the step-function, and the overdot denotes the derivative with respect to $t$.
The secular evolution of the constants of motion $(\mathcal{E}, \mathcal{L}_z, \mathcal{Q})$ under gravitational radiation reaction is calculated using the time-averaged fluxes at null infinity and the event horizon~\citep{Drasco_2004, Drasco_2006, Lo_2024} 
\begin{equation}
    \left<\dot{\E}\right> = -\mu^2 \sum_{\ell mkn} \frac{1}{4\pi\omega^2_{mkn}} \left(|\tilde{Z}^{\infty}_{\ell mkn}|^2 + \alpha_{\ell mkn}|\tilde{Z}^{\mathcal{H}}_{\ell mkn}|^2\right),
\end{equation}
\begin{equation}
    \left<\AM\right> = -\mu^2 \sum_{\ell mkn} \frac{m}{4\pi\omega^3_{mkn}} \left(|\tilde{Z}^{\infty}_{\ell mkn}|^2 + \alpha_{\ell mkn}|\tilde{Z}^{\mathcal{H}}_{\ell mkn}|^2\right),
\end{equation}
\begin{equation}
    \left<\dot{\Q}\right> = \left<\dot{\K}\right> - 2(a\E - \AM) \left(a \left<\dot{\E}\right> - \left<\dot{\AM}\right>\right),
\end{equation}
\begin{equation}
\begin{split}
    &\left<\dot{\K}\right>=2\Omega_t\left<\dot{E}\right>-\Omega_{\phi}\left<\dot{L}_z\right>\\
    &+\mu^3 \sum_{\ell mkn} \frac{n\Omega_r}{2\pi\omega^3_{mkn}} \left(|\tilde{Z}^{\infty}_{\ell mkn}|^2 + \alpha_{\ell mkn}|\tilde{Z}^{\mathcal{H}}_{\ell mkn}|^2\right),\\
\end{split}
\end{equation}
where
\begin{equation}
   \alpha_{\ell mkn}= \frac{256(2Mr_+)^5P(P^2+4\epsilon^2)(P^2+16\epsilon^2)\omega_{mkn}^3}{C^{TS}_{\ell mkn}},
\end{equation}
$P=\omega_{nkm}-\Omega_{\mathcal{H}}$ is the 
frequency shift from the horizon frequency $\Omega_{\mathcal{H}}=a/(2Mr_+)$, $\epsilon=\sqrt{M^2-a^2}/(4Mr_+)$, $r_+=M+\sqrt{M^2-a^2}$ is the outer Cauchy horizon, $C^{TS}_{\ell mkn}$ is the Teukolsky-Starobinsky constant, $\Omega_{i}=\dot{x}^i$, and $\tilde{Z}^{\mathcal{H}}_{\ell mkn}$ and $\tilde{Z}^{\infty}_{\ell mkn}$ denote the GW flux at future null infinity $\mathcal{I}^+$ and the outer Cauchy horizon $\mathcal{H}$, respectively.
To evolve the trajectory, we use the augmented analytic kludge model in the \texttt{FastEMRIWaveforms} code~\citep{Barack_2004, Babak_2007, Chua_2015, Chua_2017, Chua_2019, Stein_2020, Fujita_2020, Chua_2021, Katz_2021, Speri_2024}, and update $(\E,\AM,\Q)$ with each ODI.
\section{Orbital Evolution via Orbiter-Disk Interactions}~\label{appendix_b}

By solving the geodesic equation using the Kerr metric in Eq.~\eqref{Kerr_metric}, the equations of motion (EOM) are
\begin{equation}
    \frac{\dd t}{\dd \lambda}=\E\left[\frac{(r^2+a^2)}{\Delta}-a^2\sin^2\theta\right]-\frac{2Mar\AM}{\Delta},
\end{equation}
\begin{equation}\label{radial}
\begin{split}
    \left(\frac{\dd r}{\dd \lambda}\right)^2=V_r(r)\equiv&\left[\E(r^2+a^2)-a\AM\right]^2\\
    &-\Delta\left[r^2+(\AM-a\E)^2+\Q\right],\\
\end{split}    
\end{equation}
\begin{equation}\label{polar}
     \left(\frac{\dd \theta}{\dd \lambda}\right)^2=V_{\theta}(\theta)\equiv\Q-\AM\cot^2\theta-a^2(1-\E^2)\cos^2\theta,
\end{equation}
\begin{equation}
    \frac{\dd \phi}{\dd\lambda}=\AM\csc^2\theta-\frac{a}{\Delta}(\AM-2Ma\E),
\end{equation}
where $\lambda$ is the Mino-Carter time~\citep{Mino_2003}.
For bound orbits, one can parametrize the EOM using the radial and polar turning points.
Solving the equation $V_r=0$, yields four roots $r_1>r_2>r_3>r_4$~\citep{Schmidt_2002, Fujita_2009, van_de_Meent_2020}.
For stable eccentric orbits, $r_1=r_a$ and $r_2=r_p$ correspond to the apocenter and pericenter radius, respectively, and their expressions as a function of $(\E,\AM,\Q)$ can be found in Refs.~\citep{Fujita_2009, van_de_Meent_2020, Hughes_2019}.
From this, one can reparametrize the radial EOM using 
\begin{equation}
    p=\frac{2r_ar_p}{r_a+r_p},\hspace{10mm}e=\frac{r_a-r_p}{r_a+r_p}.
\end{equation}
Similarly, the polar by solving for the polar turning points, the polar EOM can be parametrized using 
\begin{equation}
    x_I = \mathrm{sgn}(\AM)\sin\theta_{-},
\end{equation}
$\theta_{-}<\theta_+$ are the polar turning points.
To determine the change in quasi-Keplerian coordinates due to orbiter-disk interactions, let $_i=(p,e,x_I)$ and $\mathcal{C}_i=(\E,\AM,\Q)$, so $\delta K_i=\frac{\partial K_i}{\partial \mathcal{C}_j}\delta \mathcal{C}_j$, which can be written as a Jacobian matrix
\begin{equation}
\begin{pmatrix}
    \delta p\\
    \delta e \\
    \delta x_I\\
\end{pmatrix}
=\begin{pmatrix}
    \frac{\partial p}{\partial \E}&\frac{\partial p}{\partial \AM}&\frac{\partial p}{\partial \Q}\\
    \frac{\partial e}{\partial \E}&\frac{\partial e}{\partial \AM}&\frac{\partial e}{\partial \Q}\\
    \frac{\partial x_I}{\partial \E}&\frac{\partial x_I}{\partial \AM}&\frac{\partial x_I}{\partial \Q}\\
\end{pmatrix}\begin{pmatrix}
    \delta \E\\
    \delta \AM \\
    \delta \Q\\
\end{pmatrix}.
\end{equation}
Note that this equation is underdetermined, and we need to impose two conditions to solve for $\delta{K}_i$.
The first condition we impose is $\delta r=0$, as we don't want the stellar-mass orbiter (SMO) to suddenly change radial position after each disk collision.
Therefore, $\delta p = p\cdot\delta e\cos\psi/(1+\cos\psi)$, as $r=p/(1+e\cos\psi)$. The second condition we impose is $\delta x_I=0$, because we assume that the collision angle $I$ is constant, as the disk is thin and will only damp the SMO's velocity without changing its direction.
Using the ODI constraints in the main text, we show how $(p,e,x_I)$ evolve with each successive ODI in Figure.~\ref{QPE_orbit}.
\section{Non-Discrete Modes and High-Frequency Tails due to Orbiter-Disk Interactions} 
\begin{figure}[ht!]
    \centering
    \includegraphics[width=\linewidth]{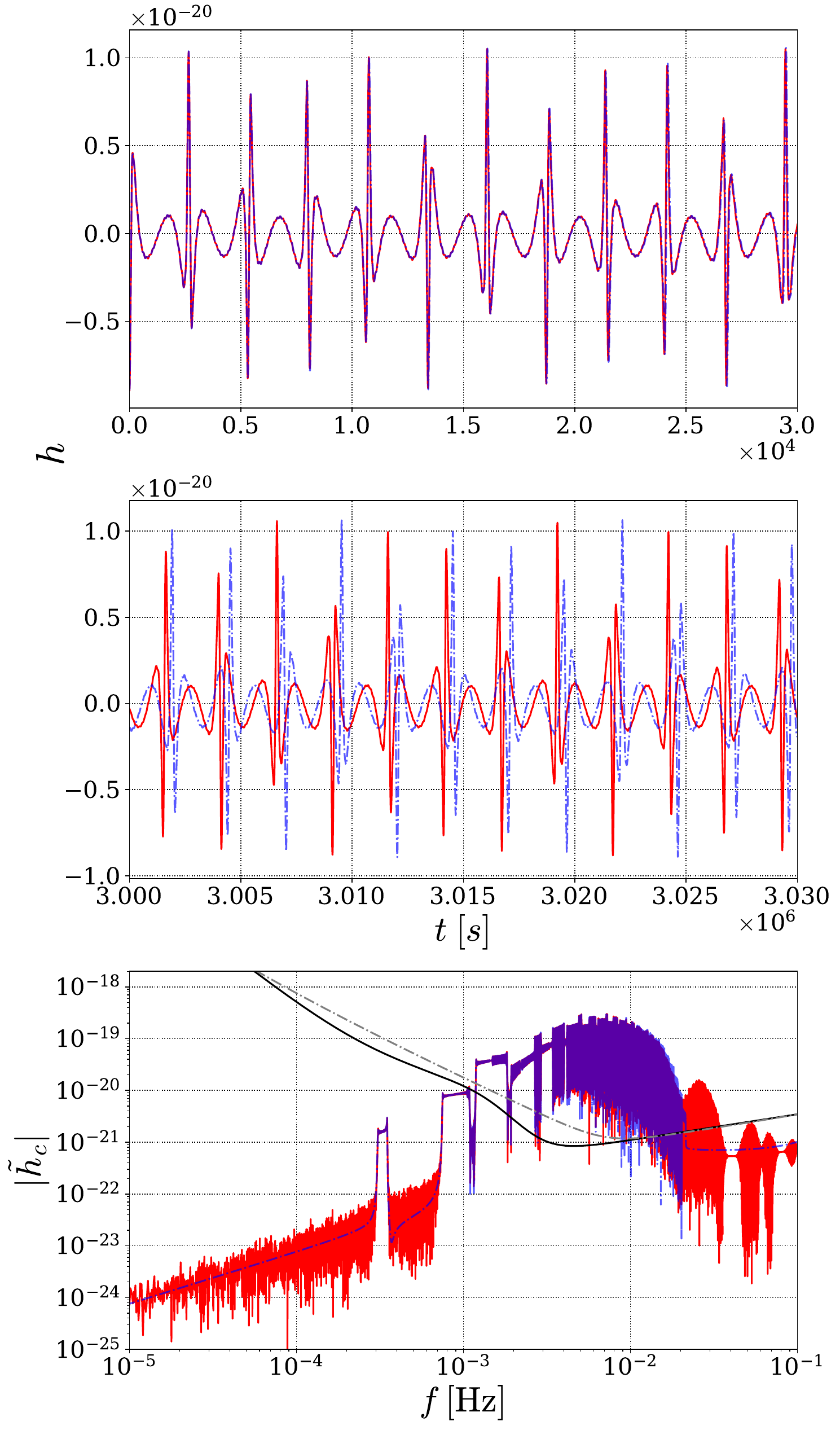}
    \caption{Snapshots of EMRI waveform at early and late times (top 2 panels), and the characteristic strain (bottom panel) comparing vacuum EMRIs (dashed blue line) versus EMRIs with ODIs (solid red line) with the noise curves of LISA (black line) and TianQin (dashed gray line).
The initial parameters are $M=10^6M_{\odot}$, $\mu = 30 M_{\odot}$, $a=0.9$, $p_0=10M$, $e_0=0.7$, $x_{I0}=\cos(\pi/4)$, $\Phi_{\phi}=\Phi_{\theta}=\Phi_{r}=0$, $q_K=\phi_K=q_S=\phi_S=0.01$, and $d_L=100\,{\rm Mpc}$.}
    \label{QPE_GW_signal}
\end{figure}
To model the gravitational waveform from an EMRI with ODIs, we consider a time-domain signal $h(t)$ experiencing impulsive changes in amplitude and frequency at collision times $t_c$
\begin{equation}
h(t) = \frac{\mu}{d_L} S(\theta,\phi) \begin{cases} 
\mathcal{A} e^{-i\Phi_{\text{pre}}(t)}, & t < t_c, \\
(\mathcal{A} + \delta\mathcal{A}) e^{-i\Phi_{\text{post}}(t)}, & t \geq t_c.
\end{cases}
\end{equation}
These discontinuities arise from sudden energy and angular momentum losses during an ODI.
Hence, the Fourier transform splits into a part with smooth segments and another part with contributions from the discontinuities.
For $t \neq t_c$, the stationary phase approximation yields quasi-monochromatic peaks $G_{\text{pre/post}}(f)$, representing the adiabatic inspiral.
The post-ODI integral,
\begin{equation}
\mathcal{I}_{\text{post}} = e^{-i2\pi f t_c} \int_{0}^{\infty} e^{-i(\delta\omega + 2\pi f)\tau} d\tau,
\end{equation}
captures the transient response to the ODI.
Using the Sokhotski-Plemelj theorem to evalute the integral for $|f| \gg |\delta\omega|$
\begin{equation}
\mathcal{I}_{\text{post}} \sim -\frac{e^{-i2\pi f t_c}}{2\pi f}.
\end{equation}

Physically, an amplitude jump $\delta\mathcal{A}$ produces dipole radiation ($1/f$ tail), while a frequency jump $\delta\omega$ creates quadrupole emission ($1/f^2$ tail) due to the discontinuous derivative in $\dot{h}(t)$, since $\mathcal{F}[\dot{h}(t)] = i2\pi f \tilde{h}(f)$.
Therefore, the frequency spectrum can be approximated as
\begin{equation}
\begin{split}
    \tilde{h}(f) \approx \frac{\mu}{d_L} S(\theta,\phi) \bigg[&\mathcal{A} G_{\text{pre}}(f)+ (\mathcal{A}+\delta\mathcal{A}) G_{\text{post}}(f)\\
     &+ \frac{\delta\mathcal{A}}{2\pi if} e^{-i\Phi(t_c)} - \frac{i\mathcal{A}\delta\omega}{(2\pi f)^2} e^{-i\Phi(t_c)} \bigg].\\
\end{split}
\end{equation}
The $G_{\text{pre/post}}$ terms represent the adiabatic inspiral before and after the ODI, producing narrow spectral peaks.
The $1/f$ term arises from instantaneous amplitude changes (dipole-like emission), while the $1/f^2$ term comes from frequency jumps that create discontinuities in the waveform derivative (quadrupole-like emission).
As shown in Figure.~\ref{QPE_GW_signal}, the cumulation of ODIs over many cycles produces a noticeable phase shift in the late inspiral due to the cumulation of small phase deviations from each ODI.
In the frequency spectrum, ODIs generate broadband spectral content between the discrete harmonic peaks of vacuum EMRIs.
The accumulated phase and amplitude deviations from multiple collisions produce high-frequency tails via constructive interference, significantly enhancing the characteristic strain $h_c(f)$ across the detector band, thus improving detectability.
\bibliography{main}{}
\bibliographystyle{aasjournalv7}

\end{document}